\documentclass[10pt,english]{article}
\usepackage{mathptmx}
\usepackage[T1]{fontenc}
\usepackage[utf8]{luainputenc}
\usepackage{babel}
\usepackage{amsmath}
\usepackage{amssymb}
\usepackage{stmaryrd}
\usepackage{graphicx}
\usepackage{setspace}
\usepackage[unicode=true,pdfusetitle,
 bookmarks=true,bookmarksnumbered=false,bookmarksopen=false,
 breaklinks=false,pdfborder={0 0 1},backref=false,colorlinks=false]
 {hyperref}

\makeatletter
\newcommand{\lyxaddress}[1]{
\par {\raggedright #1
\vspace{1.4em}
\noindent\par}
}

\makeatother

\begin{document}

\title{A simplified Parisi Ansatz}

\date{~}

\author{Simone Franchini}

\maketitle

\lyxaddress{\begin{center}
\textit{\small{}Sapienza Università di Roma, 1 Piazza Aldo Moro,
00185 Roma, Italy}
\par\end{center}}
\begin{abstract}
Based on simple combinatorial arguments, we formulate a generalized
cavity method where the Random Overlap Structure (ROSt) probability
space of Aizenmann, Sims and Starr is obtained in a constructive way,
and use it to give a simplified derivation of the Parisi formula for
the free energy of the Sherrington-Kirckpatrick model
\end{abstract}

\section{Introduction}

Some decades ago a very sophisticated mean field theory has been developed
by Parisi to compute the thermodynamic properties of the Sherrington-Kirckpatrick
(SK) model in the low temperature phase \cite{Parisi-1,PMM,Guerra,Talagarand,DotK}.
In his theory, that is obtained within the larger framework of Replica
Theory \cite{PMM,DotK}, Parisi introduced many important concepts
that are now standards of the field, like the overlap distribution
as order parameter and the nontrivial hypothesis that the scalar products
between independent replicas of the system (overlaps) concentrates
on a numeric support that is ultrametrically organized \cite{PMM,Guerra,DotK,UltraBolt,ASS,Bolt,PaK,kUltrapanchenko}. 

After many years Guerra \cite{Guerra} and Talagrand \cite{Talagarand}
showed that this remarkable mean field theory indeed provides the
correct expression for the free energy of the SK model, while Panchenko
proved that the SK Gibbs measure can be perturbed into a special cascade
of Point Processes (Ruelle Cascade \cite{Bolt,PaK}) that gives the
same free energy and indeed satisfy the ultrametricity assumption
\cite{PaK,kUltrapanchenko}. These mathematical milestones and many
other theoretical and numerical tests (see \cite{Marinero} and references)
contributed to form the idea that at least for mean-field models this
ansatz provides the correct physical properties. 

Following simple combinatorial arguments we show that the same results
of the RSB theory can be obtained in a constructive way without relying
on the replica trick, nor averaging on the disorder. After presenting
a general analysis of the SK Hamiltonian, we will show that the usual
assumptions associated to $L$ levels of Replica Symmetry Breakings
(RSB, see \cite{PMM,Mezard,ASS,Bolt,PaK}) are consistent with a hierarchical
mean field (MF) theory in which the states ensemble is charted according
to a sigma algebra generated by a partition of the vertex set. The
method provides a constructive derivation of the Random Overlap Structure
(ROSt) probability space introduced in \cite{ASS} by Aizenmann, Sims
and Starr. We further tested by computing the corresponding incremental
pressure that one obtains from the Cavity method \cite{PMM,ASS,Bolt}
and it indeed provides the correct Parisi functional.

We start by introducing the basic notation. Let consider a spin system
of $N$ spins, we indicate the spins sites by the vertex set $V=\left\{ 1,2,\,...\,,\,N\right\} $,
marked by the label $i$. To each vertex is associated a unique spin
variable $\sigma_{i}$ that can be plus or minus. Formally $\sigma_{i}\in\Omega$,
hereafter we assume $\Omega=\left\{ +,-\right\} $, although our argument
holds for any size of $|\Omega|$ (for this paper a modulus $|\,\cdot\,|$
applied to a discrete set returns its cardinality, for example $|V|=N$).
We collect the spins into the vector 
\begin{equation}
\sigma_{V}=\left\{ \sigma_{i}\in\Omega:\,i\in V\right\} 
\end{equation}
that is supported by the $N-$spin vector space $\Omega^{V}$, we
call these vectors magnetization states. Notice that we implicitly
established an arbitrary reference frame on $V$ by labeling the spins. 

Let $J$ be some matrix of entries $J_{ij}=O\left(1\right)$. Even
if the arguments we are going to present are not limited to this case,
in the following we also assume that the $J_{ij}$ entries are random
and normally distributed. Then the Sherrington-Kirckpatrick model
without external field is described by the Hamiltonian
\begin{equation}
H\left(\sigma_{V}\right)=\frac{1}{\sqrt{N}}\sum_{\left(i,j\right)\in W}\sigma_{i}J_{ij}\sigma_{j}
\end{equation}
where $W=V\varotimes V$ is the edges set accounting for the possible
spin-spin interactions and $\sqrt{N}$ is a normalization, that in
mean field models can be $|V|-$dependent. In the case of the SK model
the interactions are normally distributed and we have to take a normalization
that is the square root of the number of spins $|V|=N$, but the same
analysis can be repeated for any coupling matrix and its relative
normalization.. As usual, we can define the partition function 
\begin{equation}
Z_{N}=\sum_{\sigma\in\Omega^{V}}\exp\left[-\beta H\left(\sigma_{V}\right)\right]\label{eq:sk}
\end{equation}
and the associated Gibbs measure
\begin{equation}
\mu\left(\sigma_{V}\right)=\frac{\exp\left[-\beta H\left(\sigma_{V}\right)\right]}{Z_{N}}.
\end{equation}
The free energy density is written in term of the pressure
\begin{equation}
p=\lim_{N\rightarrow\infty}\frac{1}{N}\log Z_{N},
\end{equation}
the free energy per spin is given by $-p/\beta$. 

A variational formula for the pressure of the SK model has been found
by Parisi \cite{Parisi-1}. Following this, and after \cite{PaK,PMM,Bolt,ASS}
it has been proven that the average pressure per spin can be computed
from the relation
\begin{equation}
E\left(p\right)=\inf_{q,\lambda}\,A_{P}\left(q,\lambda\right)
\end{equation}
where $A_{P}\left(q,\lambda\right)$ is the Parisi functional of the
(asymmetric) SK model as defined in Eq. (\ref{eq:sk}), hereafter
for the noise average we use the special notation $E\left(\,\cdot\,\right)$.

The minimizer is taken over two non-decreasing sequences $q=\left\{ q_{0},q_{1},\,...\,,q_{L}\right\} $
and $\lambda=\left\{ \lambda_{0},\lambda_{1},\,...\,,\lambda_{L}\right\} $
such that $q_{0}=\lambda_{0}=0$ and $q_{L}=\lambda_{L}=1$. The Parisi
functional is defined as follows 
\begin{equation}
A_{P}\left(q,\lambda\right)=\log2+\log Y_{0}-\frac{\beta^{2}}{2}\sum_{\ell\leq L}\lambda_{\ell}\left(q_{\ell}^{2}-q_{\ell-1}^{2}\right),\label{eq:parisi}
\end{equation}
where to obtain $Y_{0}$ we apply the recursive formula $Y_{\ell-1}^{\lambda_{\ell}}=E_{\ell}\,Y_{\ell}^{\lambda_{\ell}}$
to the initial condition 
\begin{equation}
Y_{L}=\cosh\left(\beta\sum_{\ell\leq L}z_{\ell}\sqrt{2q_{\ell}-2q_{\ell-1}}\right),
\end{equation}
with $z_{\ell}$ i.i.d. normally distributed and $E_{\ell}\left(\,\cdot\,\right)$
normal average that acts on $z_{\ell}$. Notice that we are using
a definition where the temperature is rescaled by a factor $\sqrt{2}$
respect to the usual Parisi functional. This is because the Hamiltonian
$H\left(\sigma_{V}\right)$ do not represent the original SK model,
where in the coupling matrix the contribution between spins placed
on the vertex pair $\left(i,j\right)$ is counted only once, but the
so called asymmetric version, that has independent energy contributions
from both $\left(i,j\right)$ and the commuted pair $\left(j,i\right)$.
The functional for the original SK model is recovered by substituting
$\beta$ with $\beta/\sqrt{2}$.

\section{Martingale representation}

Let partition the vertex set $V$ into a number $L$ of subsets $V_{\ell}$,
with $\ell$ from $1$ to $L$. Notice that by introducing the partition
$V_{\ell}$ we are implicitly defining the invertible map that establish
which vertex $i$ is placed in which subset $V_{\ell}$, but, as we
shall see, the relevant information is in the sizes $|V_{\ell}|=N_{\ell}$
and we don't need to describe the map in detail. The partition of
$V$ induces a partition of the state 
\begin{equation}
\sigma_{V}=\left\{ \sigma_{V_{\ell}}\in\Omega^{V_{\ell}}:\,\ell\leq L\right\} 
\end{equation}
and its support. We call the sub-vectors $\sigma_{V_{\ell}}$ the
local magnetization states of $\sigma_{V}$ respect to $V_{\ell}$
, formally 
\begin{equation}
\sigma_{V_{\ell}}=\left\{ \sigma_{i}\in\Omega:\,i\in V_{\ell}\right\} .
\end{equation}

From the above definitions we can construct the sequence of vertex
sets $Q_{\ell}$ that is obtained joining the $V_{\ell}$ sets in
sequence, according to the label $\ell$
\begin{equation}
Q_{\ell}=\bigcup_{t\leq\ell}\,V_{t},
\end{equation}
this sequence is such that $Q_{\ell}\setminus Q_{\ell-1}=V_{\ell}$,
the terminal point is $Q_{L}=V$ by definition (we remark that the
order is arbitrary). Hereafter we will assume that the sets $Q_{\ell}$
are of $O$$\left(N\right)$ in cardinality, the size of each set
is given by $\left|Q_{\ell}\right|=q_{\ell}N$, the parameters are
such that $q_{L}=1$ and $q_{\ell-1}\leq q_{\ell}$. The associated
sequence of states is obtained by joining the local magnetization
states, one obtains 
\begin{equation}
\sigma_{Q_{\ell}}=\bigcup_{t\leq\ell}\,\sigma_{V_{t}}\in\Omega^{\,Q_{\ell}}
\end{equation}
composed by the first $\ell$ sub-states $\sigma_{V_{\ell}}$. Also
in this case hold the relations $\sigma_{Q_{\ell}}\setminus\sigma_{Q_{\ell-1}}=\sigma_{V_{\ell}}$
and $\sigma_{Q_{L}}=\sigma_{V}$. Notice that the sets $V_{\ell}$
are given the differences between consecutive $Q_{\ell}$ sets, then
\begin{equation}
\left|V_{\ell}\right|=\left|Q_{\ell}\right|-\left|Q_{\ell-1}\right|=\left(q_{\ell}-q_{\ell-1}\right)N.
\end{equation}

In this section we will show a martingale representation for the Gibbs
measure $\mu\left(\sigma_{V}\right)$ where we interpret the full
system as the terminal point of a sequence of subsystems of increasing
size. Formally, we show that one can split $H\left(\sigma_{V}\right)$
into a sum of ``layer Hamiltonians'' 
\begin{equation}
H\left(\sigma_{V}\right)=\sum_{\ell\leq L}H_{\ell}\left(\sigma_{Q_{\ell}}\right),
\end{equation}
each $H_{\ell}$ describing the layer of spins $V_{\ell}$ plus an
external field that account for the interface interaction with the
previous layer. 
\begin{figure}
\begin{centering}
\includegraphics[scale=0.2]{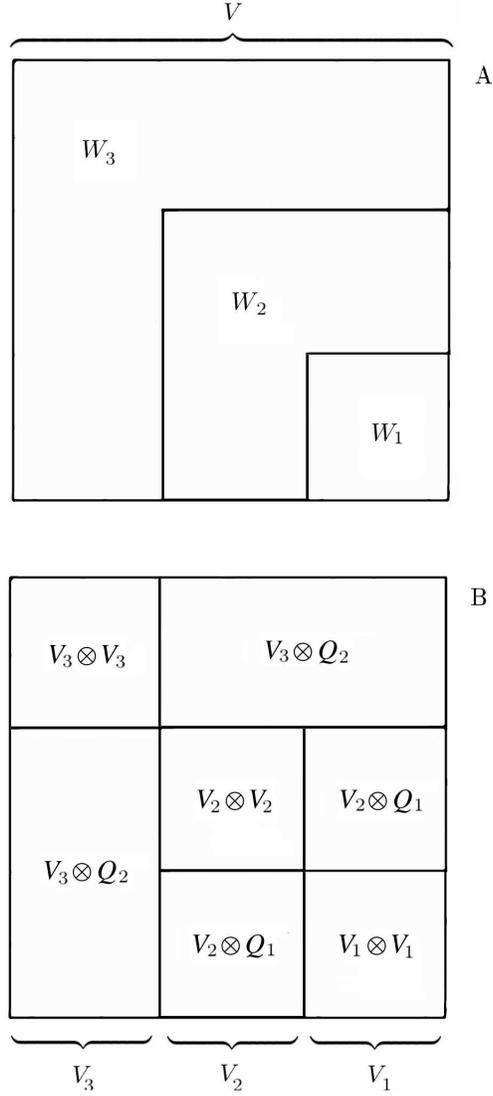}
\par\end{centering}

\raggedleft{}~\caption{\label{fig:Partition-of-}Figure A on top shows the partition of $V\varotimes V$
following that of $V$ for $L=3$. The edges set is splitted into
subsets $W_{\ell}$ containing all edges with both ends in $Q_{\ell}$
minus those with both ends in $Q_{\ell-1}$. The bottom figure B is
intended to explain the structure of $W_{\ell}$ in terms of layers
of spins: $V_{\ell}\varotimes V_{\ell}$ contain the edges between
the spins of $V_{\ell}$ while $V_{\ell}\varotimes Q_{\ell-1}$ and
$Q_{\ell-1}\varotimes V_{\ell}$ contain the edges that make the interface
between the layer $V_{\ell}$ and the rest of the system.}
\end{figure}

To prove this we first notice that the partition of the edges set
$W$ induced by that of $V$ is into subsets $W_{\ell}$ that contains
the edges with both ends in $Q_{\ell}$ minus those with both ends
in $Q_{\ell-1}$, this is also shown in the diagram of Figure \ref{fig:Partition-of-}A
where the edges $\left(i,j\right)$ are represented as points on the
square $V\varotimes V$. The Hamiltonian $H\left(\sigma_{V}\right)$
can be written as a sum of layer Hamiltonians defined as follows
\begin{equation}
H_{\ell}\left(\sigma_{Q_{\ell}}\right)=\frac{1}{\sqrt{|V|}}\sum_{\left(i,j\right)\in W_{\ell}}\sigma_{i}J_{ij}\sigma_{j},
\end{equation}
each contains the energy contributions from $W_{\ell}=\left(Q_{\ell}\varotimes Q_{\ell}\right)\setminus\left(Q_{\ell-1}\varotimes Q_{\ell-1}\right)$.
The total number of energy contributions $\sigma_{i}J_{ij}\sigma_{j}$
given by $W_{\ell}$ is 
\begin{equation}
\left|W_{\ell}\right|=|Q_{\ell}|^{2}-|Q_{\ell-1}|^{2}=\left(q_{\ell}^{2}-q_{\ell-1}^{2}\right)N^{2},
\end{equation}
that already unveils a familiar coefficient of the Parisi formula.
We can further rearrange the components of the layer contributions
by noticing that 
\begin{equation}
\left(Q_{\ell}\varotimes Q_{\ell}\right)\setminus\left(Q_{\ell-1}\varotimes Q_{\ell-1}\right)=\left(V_{\ell}\varotimes V_{\ell}\right)\cup\left(V_{\ell}\varotimes Q_{\ell-1}\right)\cup\left(Q_{\ell-1}\varotimes V_{\ell}\right),\label{eq:ssss}
\end{equation}
where the right side of the equation is also shown in Figure \ref{fig:Partition-of-}B.
Then, the energy contributions coming from $W_{\ell}$ can be rewritten
as follows
\begin{equation}
\sum_{\left(i,j\right)\in W_{\ell}}\sigma_{i}J_{ij}\sigma_{j}=\sum_{i\in V_{\ell}}\sum_{j\in V_{\ell}}\sigma_{i}J_{ij}\sigma_{j}+\sum_{i\in V_{\ell}}\sum_{j\in Q_{\ell-1}}\sigma_{i}\left(J_{ij}+J_{ji}\right)\sigma_{j}\label{eq:as}
\end{equation}
and we can identify two components, one is the layer self-interaction,
that depends only on the spins $\sigma_{V_{\ell}}$ 
\begin{equation}
\sum_{i\in V_{\ell}}\sum_{j\in V_{\ell}}\sigma_{i}J_{ij}\sigma_{j}=\sqrt{|V_{\ell}|}\,H\left(\sigma_{V_{\ell}}\right).
\end{equation}
The second contribution can be interpreted as the interface interaction
between the layers. Let define the interface fields 
\begin{equation}
h_{V_{\ell}}\left(\sigma_{Q_{\ell-1}}\right)=\left\{ h_{i}\left(\sigma_{Q_{\ell-1}}\right)\in\mathbb{R}:\,i\in V_{\ell}\right\} ,
\end{equation}
where the individual components are defined as follows 
\begin{equation}
h_{i}\left(\sigma_{Q_{\ell-1}}\right)=\frac{1}{\sqrt{|Q_{\ell-1}|}}\sum_{j\in Q_{\ell-1}}\left(J_{ij}+J_{ji}\right)\sigma_{j},
\end{equation}
then the interface contributions can be written in terms of a perturbation
depending on the preceding layers. Using these definitions into the
previous equation we find that the SK Hamiltonian can be written as
a sum of the layer energy contributions
\begin{equation}
H_{\ell}\left(\sigma_{Q_{\ell}}\right)=\sqrt{q_{\ell}-q_{\ell-1}}\,H\left(\sigma_{V_{\ell}}\right)+\sqrt{q_{\ell-1}}\,\sigma_{V_{\ell}}\cdot h_{V_{\ell}}\left(\sigma_{Q_{\ell-1}}\right).\label{eq:fgf}
\end{equation}
Notice that the contributions of the $\ell-$th level only depend
on the spins of $V_{\ell}$ and the previous $V_{t}$ for $t<\ell$,
but not on those for $t>\ell$, this is expression of the fact that
the original system is reconstructed trough an adapted process, in
which we start from the unperturbed seed $H\left(\sigma_{V_{1}}\right)$
of $N_{1}$ spins and then add layers of $N_{\ell}$ spins until reaching
the size $N$. Also, notice the coefficient $\sqrt{q_{\ell}-q_{\ell-1}}$
in front of $H\left(\sigma_{V_{\ell}}\right)$ that is due to the
$N-$dependent normalization of the SK Hamiltonian. This coefficient
is special for fully connected random models, for a fully connected
static model, like the Crurie-Weiss, would have been of order $q_{\ell}-q_{\ell-1}$,
while for models with finite connectivity the coefficient is $O\left(1\right)$,
as we shall see in short. 

From the last equations we find the corresponding partition of the
Gibbs measure. The partition function is obtained from the formula
\begin{multline}
Z_{N}=\sum_{\sigma_{V_{1}}\in\Omega^{V_{1}}}\exp\left[-\beta H_{1}\left(\sigma_{Q_{1}}\right)\right]\,...\,\sum_{\sigma_{V_{\ell}}\in\Omega^{V_{\ell}}}\exp\left[-\beta H_{\ell}\left(\sigma_{Q_{\ell}}\right)\right]\,...\,\\
\,...\,\sum_{\sigma_{V_{L}}\in\Omega^{V_{L}}}\exp\left[-\beta H_{L}\left(\sigma_{Q_{L}}\right)\right]\label{eq:sf-1}
\end{multline}
Let introduce the ``layer distributions'' 
\begin{equation}
\xi_{\ell}\left(\sigma_{Q_{\ell}}\right)=\frac{\exp\left[-\beta\,H_{\ell}\left(\sigma_{Q_{\ell}}\right)\right]}{Z_{N_{\ell}}\left(\sigma_{Q_{\ell-1}}\right)}
\end{equation}
with the layer partition functions given by
\begin{equation}
Z_{N_{\ell}}\left(\sigma_{Q_{\ell-1}}\right)=\sum_{\sigma_{V_{\ell}}\in\Omega^{V_{\ell}}}\exp\left[-\beta H_{\ell}\left(\sigma_{Q_{\ell}}\right)\right]
\end{equation}
It is easy to verify that their products gives back the original Gibbs
measure
\begin{equation}
\mu\left(\sigma_{V}\right)=\prod_{\ell\leq L}\,\xi_{\ell}\left(\sigma_{Q_{\ell}}\right),
\end{equation}
but notice that the relative weights $\xi_{\ell}\left(\sigma_{Q_{\ell}}\right)$
are measures themselves and sum to one in $\sigma_{V_{\ell}}$
\begin{equation}
\sum_{\sigma_{V_{\ell}}\in\Omega^{V_{\ell}}}\xi_{\ell}\left(\sigma_{Q_{\ell}}\right)=1,\ \forall\,\sigma_{Q_{\ell-1}}\in\Omega^{\,Q_{\ell-1}}.
\end{equation}

We can finally write the martingale representation we where searching
for. Consider the test function $f:\,\Omega^{V}\rightarrow\mathbb{R}$,
then, applying the previous definitions the average $\langle f\left(\boldsymbol{\sigma}_{V}\right)\rangle_{\mu}$
respect to $\mu$ is obtained through the following backward recursion.
The initial condition is $f_{L}\left(\sigma_{Q_{L}}\right)=f\left(\sigma_{Q_{L}}\right)$,
where $Q_{L}=V$, then we iterate the formula 

~
\begin{equation}
f_{\ell-1}\left(\sigma_{Q_{\ell-1}}\right)=\sum_{\sigma_{V_{\ell}}\in\Omega^{V_{\ell}}}\xi_{\ell}\left(\sigma_{Q_{\ell}}\right)\,f_{\ell}\left(\sigma_{Q_{\ell}}\right)\label{eq:rekurs1}
\end{equation}
backward until the first step $\ell=0$ that gives the average of
$f$ respect to the Gibbs measure $\mu$. This result is an expression
of the Bayes rule and can be easily derived starting from the identity
\begin{equation}
\mu\left(\sigma_{V}\right)=\sum_{\tau_{V}\in\Omega^{V}}\mu\left(\tau_{V}\right)\prod_{i\in V}\left({\textstyle \frac{1+\tau_{i}\sigma_{i}}{2}}\right)
\end{equation}
and substituting the definitions given before in that of $\langle f\left(\boldsymbol{\sigma}_{V}\right)\rangle_{\mu}$
brings to the desired result. Notice that up to now our manipulations
base on general principles and do not require any special assumption
concerning the Hamiltonian. 

Before going further we remark that these arguments are not limited
to mean field models, for example, we can easily extend this description
to the Ising Spin Glass in finite dimensions. 

Let $\Lambda$ be the adjacency matrix of the hyper-cubic lattice
$\mathbb{Z}^{d}$ and substitute the Hadamard product $\Lambda\circ\boldsymbol{J}$
on behalf of $\boldsymbol{J}$ and $\sqrt{g\left(\Lambda\right)}$
on behalf of $\sqrt{|V|}$, where the norm $g\left(\Lambda\right)$
is the average number of nearest-neighbors of a vertex according to
$\Lambda$, 
\begin{equation}
g_{V}\left(\Lambda\right)=\frac{1}{|V|}\sum_{i\in V}\sum_{j\in V}I\left(|\Lambda_{ij}|>0\right).
\end{equation}
If the adjacency matrix $\Lambda$ is that of a fully connected graph
we take $g\left(\Lambda\right)=|V|$ and recover the ASK model, otherwise
for $\mathbb{Z}^{d}$ is $g\left(\Lambda\right)=2d$. The result is
the following generalized Hamiltonian 
\begin{equation}
H_{\Lambda}\left(\sigma_{V}\right)=\frac{1}{\sqrt{g_{V}\left(\Lambda\right)}}\sum_{i\in V}\sum_{j\in V}\sigma_{i}\Lambda_{ij}J_{ij}\sigma_{j}
\end{equation}
If the adjacency matrix is fully connected, which is the case of SK
and other mean field models, there is no underlying geometry associated
to $V$ and we can grow the system the size we want. In finite dimensional
models, however, we may have additional constraints. In the finite
dimensional case, to grow an Ising spin glass on $\mathbb{Z}^{d}$
we should consider a cube that is enclosed in a larger cube and so
on. To enclose an hyper-cubic region of $\mathbb{Z}^{d}$ of side
lenght $r$ and volume $r^{d}$ into a larger region of side $r+k$
we need at least $\left(r+k\right)^{d}-r^{d}$ new sites to add, so
the sizes of the $V$ partition should satisfy the relation $|Q_{\ell}|=r_{\ell}^{d}$,
or equivalently $|V_{\ell}|=r_{\ell}^{d}-r_{\ell-1}^{d},$ for some
integer sequence $r_{\ell}$. 

Due to the presence of $g_{V}\left(\Lambda\right)$ nearest neighbors
to each site, each layer contributes to the total energy with $|W_{\ell}|=g_{V}\left(\Lambda\right)|V_{\ell}|$
edges, each multiplied by its coupling $J_{ij}$. Apart from this,
the partition works in the same way
\begin{equation}
H_{\ell}\left(\sigma_{Q_{\ell-1}},\sigma_{V_{\ell}}\right)=H_{\Lambda_{\ell}}\left(\sigma_{V_{\ell}}\right)+\sigma_{V_{\ell}}h_{V_{\ell}}\left(\sigma_{Q_{\ell-1}}\right),\label{MARGINALSSSSSHAM-1}
\end{equation}
where the local contributions are defined as follows
\begin{equation}
H_{\Lambda_{\ell}}\left(\sigma_{V_{\ell}}\right)=\frac{1}{\sqrt{g_{V}\left(\Lambda\right)}}\sum_{i\in V_{\ell}}\sum_{j\in V_{\ell}}\sigma_{i}\Lambda_{ij}J_{ij}\sigma_{j}
\end{equation}
and the cavity fields again incorporate the interface interaction
between the layers
\begin{equation}
h_{i}\left(\sigma_{Q_{\ell-1}}\right)=\frac{1}{\sqrt{g_{V}\left(\Lambda\right)}}\sum_{j\in Q_{\ell-1}}\left(\Lambda_{ij}J_{ij}+\Lambda_{ji}J_{ji}\right)\sigma_{j}.
\end{equation}
For this paper we concentrate on the mean field description.

\section{Incremental pressure}

To make the previous formulas effective we need a way to express the
pressure in terms of the Gibbs measure. This can be done by the Cavity
Method \cite{PMM,Mezard,ASS}, ie by relating the partition function
of an $N-$spin system with that of a larger $(N+1)-$system and then
computing the difference between the logarithms of the partition functions. 

In this paper we follow a derivation in \cite{UltraBolt} originally
due to Aizenmann et al. \cite{ASS}, see also \cite{Bolt,PaK}. Define
the cavity variables, ie the cavity field
\begin{equation}
\tilde{x}\left(\sigma_{V}\right)=\sqrt{\frac{2}{N}}\sum_{i\in V}\tilde{J}_{ii}\sigma_{i}
\end{equation}
and the so called ``fugacity term'' (see\cite{ASS})
\begin{equation}
\tilde{y}\left(\sigma_{V}\right)=\frac{1}{N}\sum_{i\in V}\sum_{j\in V}\sigma_{i}\tilde{J}_{ij}\sigma_{j}=\frac{1}{\sqrt{N}}\tilde{H}\left(\sigma_{V}\right)
\end{equation}
that is proportional to the Hamiltonian in distribution, with a different
noise matrix. First we apply the Gaussian summation rule
\begin{equation}
J_{ij}/\sqrt{N}\stackrel{d}{=}J_{ij}/\sqrt{N+1}+\tilde{J}_{ij}/\sqrt{N\left(N+1\right)}
\end{equation}
to the Hamiltonian of the $N-$system to isolate the fugacity term.
The matrix $\tilde{J}$ is a new noise independent from the $J$.
The following relation holds in distribution
\begin{multline}
H{\textstyle \left(\sigma_{V}\right)}=\frac{1}{\sqrt{N}}\sum_{i\in V}\sum_{j\in V}\sigma_{i}J_{ij}\sigma_{j}\stackrel{d}{=}\\
\stackrel{d}{=}\frac{1}{\sqrt{N+1}}\sum_{i\in V}\sum_{j\in V}\sigma_{i}J_{ij}\sigma_{j}+\frac{1}{\sqrt{N\left(N+1\right)}}\sum_{i\in V}\sum_{j\in V}\sigma_{i}\tilde{J}_{ij}\sigma_{j},\label{eq:sf}
\end{multline}
using the definition of $\tilde{y}\left(\sigma_{V}\right)$ the partition
function is written as
\begin{equation}
Z_{N}\stackrel{d}{=}\sum_{\sigma_{V}\in\Omega^{V}}\exp{\textstyle \left(-\beta\sqrt{\frac{N}{N+1}}H{\textstyle \left(\sigma_{V}\right)}\right)}\cdot\exp{\textstyle \left(\beta\sqrt{\frac{N}{N+1}}\,\tilde{y}\left(\sigma_{V}\right)\right)}\label{eq:sdg}
\end{equation}
notice that the average is respect to a $N-$system at slightly shifted
temperature. Now consider the system of $N+1$ spins, we separate
the last spin to find
\begin{multline}
H{\textstyle \left(\sigma_{V\cup\,\left\{ N+1\right\} }\right)}=\frac{1}{\sqrt{N+1}}\sum_{i\in V\cup\,\left\{ N+1\right\} }\sum_{j\in V\cup\,\left\{ N+1\right\} }\sigma_{i}J_{ij}\sigma_{j}=\\
=\frac{1}{\sqrt{N+1}}\sum_{i\in V}\sum_{j\in V}\sigma_{i}J_{ij}\sigma_{j}+\frac{1}{\sqrt{N+1}}\,\sigma_{N+1}\sum_{i\in V}\left(J_{i,N+1}+J_{N+1,i}\right)\sigma_{i}+\\
+O{\textstyle \left(\frac{1}{\sqrt{N+1}}\right)}.\label{eq:ddddd}
\end{multline}
Since the sequence $J_{i,N+1}$ and its transposed are independent
from the other $J$ entries and also between themselves, we can write
a more pleasant formula by using the diagonal terms of $\tilde{J}$
on behalf, ie we use again the Gaussian summation rule 
\begin{equation}
J_{i,N+1}+J_{N+1,i}\stackrel{d}{=}\tilde{J}_{ii}\sqrt{2},
\end{equation}
where the superscript $d$ specify that the equality holds in distribution.
The noise relative to the vertex $N+1$ is written entirely in terms
of the $\tilde{J}$ matrix. The associated partition function is computed
by integrating the spin $\sigma_{N+1}$, one obtains
\begin{equation}
Z_{N+1}\stackrel{d}{=}\sum_{\sigma_{V}\in\Omega^{V}}\exp{\textstyle \left(-\beta\sqrt{\frac{N}{N+1}}H{\textstyle \left(\sigma_{V}\right)}\right)}\cdot2\cosh\left({\textstyle \beta\sqrt{\frac{N}{N+1}}\,\tilde{x}\left(\sigma_{V}\right)}\right)\label{eq:sdg-1}
\end{equation}
Now both partition functions are rewritten in terms of the $N-$system
at rescaled temperature 
\begin{equation}
\beta^{*}=\beta\sqrt{N/\left(N+1\right)}.
\end{equation}
We distinguish the rescaled partition function from $Z_{N}$ with
a star in superscript
\begin{equation}
Z_{N}^{*}=\sum_{\sigma_{V}\in\Omega^{V}}\exp{\textstyle \left[-\beta^{*}H{\textstyle \left(\sigma_{V}\right)}\right]}.
\end{equation}
Dividing by $Z_{N}^{*}$ both $Z_{N+1}$ and $Z_{N}$ we can eventually
write the incremental pressure in terms of the measure
\begin{multline}
\log\,Z_{N+1}-\log Z_{N}\stackrel{d}{=}\\
=\log\sum_{\sigma_{V}\in\Omega^{V}}\frac{\exp{\textstyle \left[-\beta^{*}H{\textstyle \left(\sigma_{V}\right)}\right]}}{Z_{N}^{*}}2\cosh\left({\textstyle \beta^{*}\,\tilde{x}\left(\sigma_{V}\right)}\right)+\\
-\log\sum_{\sigma_{V}\in\Omega^{V}}\frac{\exp{\textstyle \left[-\beta^{*}H{\textstyle \left(\sigma_{V}\right)}\right]}}{Z_{N}^{*}}\exp{\textstyle \left(\beta^{*}\,\tilde{y}\left(\sigma_{V}\right)\right)}=\\
=\log\sum_{\sigma_{V}\in\Omega^{V}}\mu^{*}\left(\sigma_{V}\right)2\cosh\left({\textstyle \beta^{*}\tilde{x}\left(\sigma_{V}\right)}\right)+\\
-\log\sum_{\sigma_{V}\in\Omega^{V}}\mu^{*}\left(\sigma_{V}\right)\exp{\textstyle \left(\beta^{*}\tilde{y}\left(\sigma_{V}\right)\right)}.\label{eq:667}
\end{multline}
Then, apart from a rescaling $\beta^{*}\rightarrow\beta$ and other
terms that are negligible in the thermodynamic limit the pressure
can be bounded from below by the incremental pressure functional,
\begin{equation}
A\left(\tilde{x},\tilde{y},\mu\right)=\log\langle2\cosh\left({\textstyle \beta\tilde{x}\left(\sigma_{V}\right)}\right)\rangle_{\mu}-\log\langle\exp{\textstyle \left(\beta\tilde{y}\left(\sigma_{V}\right)\right)}\rangle_{\mu}.\label{eq:ghf}
\end{equation}
because the pressure is always bounded from below by the limit inferior
of the incremental pressure
\begin{equation}
p\geq\liminf_{N\rightarrow\infty}\ \log\frac{Z_{N+1}}{Z_{N}}\stackrel{d}{=}\liminf_{N\rightarrow\infty}\ A\left(\tilde{x},\tilde{y},\mu\right).
\end{equation}
Until this point the analysis is well known. Let now apply some considerations
from the previous section to the cavity variables. The cavity field
is easy, as it is natural to split
\begin{equation}
{\textstyle \tilde{x}\left(\sigma_{V}\right)}=\sqrt{\frac{2}{N}}\sum_{i\in V}\tilde{J}_{ii}\sigma_{i}=\sqrt{\frac{2}{N}}\sum_{\ell\leq L}\tilde{z}_{\ell}\left(\sigma_{V_{\ell}}\right)\sqrt{\left|V_{\ell}\right|}
\end{equation}
into independent variables that are functions of the $V_{\ell}$ spins
only
\begin{equation}
\tilde{z}_{\ell}\left(\sigma_{V_{\ell}}\right)\sqrt{\left|V_{\ell}\right|}=\sum_{i\in V_{\ell}}\tilde{J}_{ii}\sigma_{i}
\end{equation}
The fugacity term is distributed like the Hamiltonian, and then we
can use the same arguments before and write the decomposition
\begin{equation}
\tilde{y}\left(\sigma_{V}\right)=\frac{1}{N}\sum_{i\in V}\sum_{j\in V}\sigma_{i}\tilde{J}_{ij}\sigma_{j}=\frac{1}{N}\sum_{\ell\leq L}\sum_{\left(i,j\right)\in W_{\ell}}\sigma_{i}\tilde{J}_{ij}\sigma_{j}=\frac{1}{N}\sum_{\ell\leq L}\tilde{g}_{\ell}\left(\sigma_{Q_{\ell}}\right)\sqrt{\left|W_{\ell}\right|}
\end{equation}
where we introduced the new variable
\begin{equation}
\tilde{g}_{\ell}\left(\sigma_{Q_{\ell}}\right)\sqrt{\left|W_{\ell}\right|}=\sum_{\left(i,j\right)\in W_{\ell}}\sigma_{i}\tilde{J}_{ij}\sigma_{j},
\end{equation}
Notice that both $\tilde{z}_{\ell}\left(\sigma_{V_{\ell}}\right)$
and $\tilde{g}_{\ell}\left(\sigma_{Q_{\ell}}\right)$ are normally
distributed respect to $\sigma_{Q_{\ell}}$, ie Gaussian instances
and of unitary variance for all $\ell$. In terms of these new variables
the old cavity variables are
\begin{equation}
\tilde{x}\left(\sigma_{V}\right)=\sum_{\ell\leq L}\tilde{z}_{\ell}\left(\sigma_{V_{\ell}}\right)\sqrt{2q_{\ell}-2q_{\ell-1}},
\end{equation}
\begin{equation}
\tilde{y}\left(\sigma_{V}\right)=\sum_{\ell\leq L}\tilde{g}_{\ell}\left(\sigma_{Q_{\ell}}\right)\sqrt{q_{\ell}^{2}-q_{\ell-1}^{2}}.
\end{equation}
and match that of the Random Overap Structure (ROSt) probability space
first introduced in \cite{ASS}. Indeed, this is precisely the point
where the martingale representation before plays its role, as it allows
to bridge between the Pure State distributions described in \cite{PMM},
that we can identify with the following products of layer distributions
\begin{equation}
\mu_{\ell}\left(\sigma_{Q_{\ell}}\right)=\prod_{k\le\ell}\xi_{k}\left(\sigma_{Q_{k}}\right),
\end{equation}
and the ROSt probability space given in \cite{ASS}, with all its
remarkable mathematical features. 

Putting together the functional becomes
\begin{multline}
A\left(q,\tilde{z},\tilde{g},\xi\right)\stackrel{d}{=}\log\,{\textstyle \left\langle \,...\,\left\langle 2\cosh\left(\beta\sum_{\ell}\tilde{z}_{\ell}\left(\sigma_{V_{\ell}}\right)\sqrt{2q_{\ell}-2q_{\ell-1}}\right)\right\rangle _{\xi_{L}}\,...\,\right\rangle _{\xi_{1}}}+\\
-\log\,{\textstyle \left\langle \,...\,\left\langle \exp\left(\beta\sum_{\ell}\tilde{g}_{\ell}\left(\sigma_{Q_{\ell}}\right)\sqrt{q_{\ell}^{2}-q_{\ell-1}^{2}}\right)\right\rangle _{\xi_{L}}\,...\,\right\rangle _{\xi_{1}}.}\label{eq:ss}
\end{multline}
In computing the previous formula we made the natural assumption that
the partition used to split the Hamiltonian $H\left(\sigma_{V}\right)$
should be the same used to split the terms that appear in the cavity
formula, then the dependence of $A$ on $q$ is both explicit and
trough the distributions $\xi_{\ell}$. It only remains to discuss
the averaging properties of the layer distributions.

\section{Simplified ansatz}

We start by noticing that due to the vanishing coefficient $\sqrt{q_{\ell}-q_{\ell-1}}$
in front of $H\left(\sigma_{V_{\ell}}\right)$ this contribution in
Eq. (\ref{eq:fgf}) can be actually neglected in the $L\rightarrow\infty$
limit. If we introduce the rescaled temperature parameter 
\begin{equation}
\beta_{\ell}=\beta\sqrt{q_{\ell}-q_{\ell-1}},
\end{equation}
that can be made arbitrarily small in the $L\rightarrow\infty$ limit,
then we can rewrite each layer in terms of an SK model of size $N_{\ell}$
at temperature $\beta_{\ell}$ 
\begin{equation}
\beta H_{\ell}\left(\sigma_{Q_{\ell}}\right)\stackrel{d}{=}\beta_{\ell}\left[H\left(\sigma_{V_{\ell}}\right)+\sigma_{V_{\ell}}\cdot h_{V_{\ell}}^{*}\left(\sigma_{Q_{\ell-1}}\right)\right]
\end{equation}
subject to the (strong) external field 
\begin{equation}
h_{V_{\ell}}^{*}\left(\sigma_{Q_{\ell-1}}\right)=\frac{1}{\sqrt{|V_{\ell}|}}\sum_{j\in Q_{\ell-1}}\left(J_{ij}+J_{ji}\right)\sigma_{j},
\end{equation}
whose magnitude diverges in the $L\rightarrow\infty$ limit due to
the $\sqrt{|V_{\ell}|}$ normalization. Then, for any finite temperature
$\beta$ we can make $N$ and $L$ large enough to have a $q_{\ell}$
sequence for which $\beta_{\ell}<\beta_{c}$ at any $\ell$, and it
is established since \cite{Talagrand MF} and \cite{AT-line} that
in the high temperature regime the annealed averages needed to compute
Eq. (\ref{eq:ss}) matches the quenched ones (the layers are Replica
Symmetric).

To make this argument more precise let consider the Hamiltonian 
\begin{equation}
\bar{H}_{\ell}\left(\sigma_{Q_{\ell}}\right)=\sqrt{q_{\ell-1}}\,\sigma_{V_{\ell}}\cdot h_{V_{\ell}}\left(\sigma_{Q_{\ell-1}}\right),\label{eq:fgf-1}
\end{equation}
in the Thermodynamic Limit and for $L\rightarrow\infty$ one can compute
the averages in Eq.(\ref{eq:rekurs1}) according to the Hamiltonian
$\bar{H}_{\ell}\left(\sigma_{Q_{\ell}}\right)$ instead of $H_{\ell}\left(\sigma_{Q_{\ell}}\right)$,
this will be shown at the end of this section. The partition function
of the $\bar{H}_{\ell}$ model can be computed exactly and one finds
\begin{multline}
\bar{Z}_{N_{\ell}}\left(\sigma_{Q_{\ell-1}}\right)=\sum_{\sigma_{V_{\ell}}\in\Omega^{V_{\ell}}}\exp\left[\beta\sqrt{q_{\ell-1}}\,\sigma_{V_{\ell}}\cdot h_{V_{\ell}}\left(\sigma_{Q_{\ell-1}}\right)\right]=\\
=\prod_{i\in V_{\ell}}2\cosh\left(\beta\sqrt{q_{\ell-1}}\,h_{i}\left(\sigma_{Q_{\ell-1}}\right)\right)=\exp\left[-\beta\bar{F}_{N_{\ell}}\left(\sigma_{Q_{\ell-1}}\right)\right].\label{eq:ujyjf}
\end{multline}
Moreover, following \cite{PMM}, from Boltzmann theory one can show
that at equilibrium the logarithm of the associated Gibbs distribution
is proportional to the fluctuations around the average internal energy
\begin{equation}
\bar{\xi}_{\ell}\left(\sigma_{Q_{\ell}}\right)\propto\exp\left[-\beta\Delta\bar{H}\left(\sigma_{Q_{\ell}}\right)\right]
\end{equation}
where the fluctuations are defined as follows 
\begin{equation}
\Delta\bar{H}\left(\sigma_{Q_{\ell}}\right)=\sqrt{q_{\ell-1}}\,\left[\sigma_{V_{\ell}}\cdot h_{V_{\ell}}\left(\sigma_{Q_{\ell-1}}\right)-\langle\sigma_{V_{\ell}}\cdot h_{V_{\ell}}\left(\sigma_{Q_{\ell-1}}\right)\rangle_{\bar{\xi}_{\ell}}\right].
\end{equation}
Notice that for the $\bar{H}_{\ell}$ model the energy overlap $\langle\Delta\bar{H}\left(\sigma_{Q_{\ell}}\right)\Delta\bar{H}\left(\tau_{Q_{\ell}}\right)\rangle_{\bar{\mu}\otimes\bar{\mu}}$
can be computed exactly, but this long algebraic work is not necessary
in order to compute the Parisi functional. In fact, by Central Limit
Theorem in the large $N$ limit the fluctuations $\Delta\bar{H}\left(\sigma_{Q_{\ell}}\right)$
converge to a Gaussian set indexed by $\sigma_{Q_{\ell-1}}$, whose
canonical variance is 
\begin{equation}
\langle\Delta H\left(\sigma_{Q_{\ell}}\right)^{2}\rangle_{\bar{\xi}_{\ell}}=\frac{\partial}{\partial\beta^{2}}\bar{F}_{N_{\ell}}\left(\sigma_{Q_{\ell-1}}\right)=N\bar{\gamma}_{\ell}\left(\sigma_{Q_{\ell-1}}\right)^{2}.
\end{equation}
Then, under the Gibbs measure $\bar{\xi}_{\ell}$ the energy fluctuations
can be approximated in distribution by a Derrida's Random Energy Model
(REM, see \cite{PaK,UltraBolt}) 
\begin{equation}
H_{\ell}\left(\sigma_{Q_{\ell}}\right)=\epsilon_{\ell}\left(N\right)+\Delta\bar{H}\left(\sigma_{V_{\ell}}\right)\stackrel{d}{=}\epsilon_{\ell}\left(N\right)+\bar{\gamma}_{\ell}\left(\sigma_{Q_{\ell-1}}\right)g_{\ell}^{*}\left(\sigma_{V_{\ell}}\right)\sqrt{N}\label{eq:approc}
\end{equation}
where $g_{\ell}^{*}\left(\sigma_{V_{\ell}}\right)$ are i.i.d. normally
distributed variables with covariance matrix
\begin{equation}
E\left[g_{\ell}^{*}\left(\sigma_{V_{\ell}}\right)g_{\ell}^{*}\left(\tau_{V_{\ell}}\right)\right]=\prod_{i\in V_{\ell}}I\left(\sigma_{i}=\tau_{i}\right).
\end{equation}
Since the SK measure is weakly exchangeable, although $\bar{\gamma}_{\ell}\left(\sigma_{Q_{\ell-1}}\right)$
may depends on the spins of $\sigma_{Q_{\ell-1}}$ trough the cavity
fields $h_{V_{\ell}}\left(\sigma_{Q_{\ell-1}}\right)$ the only way
to enforce this invariance is to admit that eventually 
\begin{equation}
\bar{\gamma}_{\ell}\left(\sigma_{Q_{\ell-1}}\right)^{2}\stackrel{d}{=}\bar{\gamma}_{\ell}^{2}
\end{equation}
under $\xi_{\ell}$ average for some positive number $\bar{\gamma}_{\ell}^{2}$.
Notice that $\bar{\gamma}_{\ell}\left(\sigma_{Q_{\ell-1}}\right)^{2}=\bar{\gamma}_{\ell}^{2}$
independent of $\sigma_{Q_{\ell-1}}$doesn't mean that the sign of
$\bar{\gamma}_{\ell}\left(\sigma_{Q_{\ell-1}}\right)$ is fixed, and
under the full measure $\bar{\mu}$ one may have different correlations
between the full states $\sigma_{Q_{\ell}}$ due to the averaging
effect of $\bar{\mu}$ on the interface fields. The term $\epsilon_{\ell}\left(N\right)$
in Eq. (\ref{eq:approc}) is a constant that does not depend on the
spins and we can interpret it as the deterministic component of $H_{\ell}\left(\sigma_{Q_{\ell}}\right)$
under Gibbs measure, for the SK model we expect $\epsilon_{\ell}\left(N\right)=0$
for all $\ell$ but its exact value is not important in computing
the Parisi functional because in the end it will washed out by the
difference between the logarithms. 

Before discussing the physical features let verify that the simplified
ansatz provides the correct Parisi functional. As is shown in \cite{Bolt},
the thermodynamic limit of a Gaussian REM of amplitude $\bar{\gamma}$
is proportional in distribution to a Poisson Point Process (PPP) of
rate 
\begin{equation}
\lambda_{\ell}=\frac{\sqrt{2\log2}}{\bar{\gamma}_{\ell}}.
\end{equation}

The system at equilibrium is then decomposed into a large (eventually
infinite) number $L$ of subsystems, one for each vertex set $V_{\ell}$,
whose Gibbs measure are proportional in distribution to a sequence
of Poisson-Dirichlet (PD) point processes, ie the Gibbs measures that
describe the layers are proportional in distribution to Poisson Point
Processes (PPP) \cite{Bolt,PaK} of rate $\lambda_{\ell}$, that is
a function of $q$ but independent from the spins $\sigma_{Q_{\ell-1}}$. 

By the special average property of PPP \cite{Bolt,PaK} (see also
the Little Theorem of \cite{Mezard}) for any positive test function
$f:\Omega^{N}\rightarrow\mathbb{R}^{+}$ we have
\begin{equation}
\sum_{\sigma_{V_{\ell}}\in\Omega^{V_{\ell}}}\xi_{\ell}\left(\sigma_{Q_{\ell}}\right)f\left(\sigma_{Q_{\ell}}\right)\stackrel{d}{=}C_{\ell}\left(\sum_{\sigma_{V_{\ell}}\in\Omega^{V_{\ell}}}f\left(\sigma_{Q_{\ell}}\right)^{\lambda_{\ell}}\right)^{1/\lambda_{\ell}}
\end{equation}
for some constant $C_{\ell}$ that may depend on $\beta$ but not
on the spins. Then the random average $\langle f\left(\boldsymbol{\sigma}_{V}\right)\rangle_{\mu}$
is obtained through the following recursion 
\begin{equation}
f_{\ell-1}\left(\sigma_{Q_{\ell-1}}\right)^{\lambda_{\ell}}\stackrel{d}{=}K_{\ell}\left(\frac{1}{2^{|V_{\ell}|}}\sum_{\sigma_{V_{\ell}}\in\Omega^{U_{\ell}}}\,f_{\ell}\left(\sigma_{V_{\ell}}\right)^{\,\lambda_{\ell}}\right),\label{eq:recursin-2}
\end{equation}
that holds in distribution, with $K_{\ell}=2^{|V_{\ell}|}\,C_{\ell}^{\lambda_{\ell}}$.
This allows to compute the main contribution
\begin{equation}
{\textstyle \left\langle \,...\,\left\langle 2\cosh\left(\beta\sum_{\ell}\tilde{z}_{\ell}\left(\sigma_{V_{\ell}}\right)\sqrt{2q_{\ell}-2q_{\ell-1}}\right)\right\rangle _{\xi_{L}}\,...\,\right\rangle _{\xi_{1}}}\stackrel{d}{=}Y_{0}\exp\left(\sum_{\ell\leq L}\log K_{\ell}\right)\label{eq:ddds}
\end{equation}
by applying the recursive relation
\begin{equation}
Y_{\ell-1}^{\lambda_{\ell}}=\frac{1}{2^{|V_{\ell}|}}\sum_{\sigma_{V_{\ell}}\in\Omega^{V_{\ell}}}Y_{\ell}^{\lambda_{\ell}}
\end{equation}
to the initial condition 
\begin{equation}
Y_{L}=2\cosh\left(\beta\sum_{\ell\leq L}\tilde{z}_{\ell}\left(\sigma_{V_{\ell}}\right)\sqrt{2q_{\ell}-2q_{\ell-1}}\right)
\end{equation}
down to the last $\ell=0$. Notice that in the recursion the average
over $\sigma_{V_{\ell}}$ is uniform, under uniform distribution both
$\tilde{z}_{\ell}\left(\sigma_{V_{\ell}}\right)$ and $\tilde{g}_{\ell}\left(\sigma_{Q_{\ell}}\right)$
are normally distributed and independent from the previous spin layers
$\sigma_{Q_{\ell-1}}$, then we can take
\begin{equation}
\tilde{z}_{\ell}\left(\sigma_{V_{\ell}}\right)\stackrel{d}{=}z_{\ell},\ \tilde{g}_{\ell}\left(\sigma_{Q_{\ell}}\right)\stackrel{d}{=}g_{\ell},
\end{equation}
with $z_{\ell}$ and $g_{\ell}$ i.i.d. normally distributed, and
change the uniform average over $\sigma_{V_{\ell}}$ into a Gaussian
average $E_{\ell}$ acting on these new variables. We compute the
fugacity term in the same way, 
\begin{multline}
{\textstyle \left\langle \,...\,\left\langle \exp\left(\beta\sum_{\ell}\tilde{g}_{\ell}\left(\sigma_{Q_{\ell}}\right)\sqrt{q_{\ell}^{2}-q_{\ell-1}^{2}}\right)\right\rangle _{\xi_{L}}\,...\,\right\rangle _{\xi_{1}}}\stackrel{d}{=}\\
\stackrel{d}{=}\exp\left(\frac{\beta^{2}}{2}\sum_{\ell\leq L}\lambda_{\ell}\left(q_{\ell}^{2}-q_{\ell-1}^{2}\right)+\sum_{\ell\leq L}\log K_{\ell}\right).\label{eq:vv}
\end{multline}
Putting together the contributions depending from $K_{\ell}$ cancel
out and one finds

\begin{equation}
\exp A_{P}\left(q,\lambda\right)=Y_{0}\exp\left(-\frac{\beta^{2}}{2}\sum_{\ell\leq L}\lambda_{\ell}\left(q_{\ell}^{2}-q_{\ell-1}^{2}\right)\right)
\end{equation}
that is exactly the Parisi functional as is defined in the introduction.
Notice that in this equation and in the previous we implicitly assumed
that the sequences $q$ and $\lambda$ are exactly those that approximate
the SK model. The lower bound in the variational formula can be easily
obtained from the knowledge of the Parisi functional by minimizing
on the possible sequences $q$ and $\lambda$ 
\begin{equation}
A\left(q,\tilde{z},\tilde{g},\xi\right)\geq\inf_{q,\lambda}\ A_{P}\left(q,\lambda\right),
\end{equation}
while the upper bound can be checked, at least for the SK model, by
Guerra-Toninelli interpolation \cite{Guerra}. 

It remain to prove that one can compute the averages in Eq.(\ref{eq:rekurs1})
according to the Hamiltonian $\bar{H}_{\ell}\left(\sigma_{Q_{\ell}}\right)$
instead of $H_{\ell}\left(\sigma_{Q_{\ell}}\right)$, consider the
full layer distribution 
\begin{equation}
\xi_{\ell}\left(\sigma_{Q_{\ell-1}}\right)=\frac{1}{Z_{N_{\ell}}\left(\sigma_{Q_{\ell-1}}\right)}\exp\left[-\beta\,H_{\ell}\left(\sigma_{Q_{\ell}}\right)\right],
\end{equation}
define its version without external field
\begin{equation}
\xi_{\ell}^{*}\left(\sigma_{V_{\ell}}\right)=\frac{1}{Z_{N_{\ell}}^{*}}\exp\left[-\beta_{\ell}\,H\left(\sigma_{V_{\ell}}\right)\right]
\end{equation}
that is simply an SK model at (eventually high) temperature $\beta_{\ell}=\beta\sqrt{q_{\ell}-q_{\ell-1}}$.
Now, if we assume the thermodynamic limit exists we can use the Boltzmann
theory and express the thermodynamics fluctuations as they where a
Gaussian set. Start from the general average formula
\begin{equation}
f_{\ell-1}\left(\sigma_{Q_{\ell-1}}\right)=\sum_{\sigma_{V_{\ell}}\in\Omega^{V_{\ell}}}\xi_{\ell}\left(\sigma_{Q_{\ell}}\right)\,f_{\ell}\left(\sigma_{Q_{\ell}}\right),
\end{equation}
using the REM-PPP relation on $H\left(\sigma_{V_{\ell}}\right)$ and
the PPP average properties one finds that the formula for the partition
function is
\begin{multline}
Z_{N_{\ell}}\left(\sigma_{Q_{\ell-1}}\right)=\sum_{\sigma_{V_{\ell}}\in\Omega^{V_{\ell}}}\exp\left[-\beta\,H_{\ell}\left(\sigma_{Q_{\ell}}\right)\right]=\\
=\sum_{\sigma_{V_{\ell}}\in\Omega^{V_{\ell}}}\exp\left[-\beta_{\ell}\,H\left(\sigma_{V_{\ell}}\right)-\beta\bar{H}_{\ell}\left(\sigma_{Q_{\ell}}\right)\right]=\\
=Z_{N_{\ell}}^{*}\sum_{\sigma_{V_{\ell}}\in\Omega^{V_{\ell}}}\xi_{\ell}^{*}\left(\sigma_{V_{\ell}}\right)\exp\left[-\beta\bar{H}_{\ell}\left(\sigma_{Q_{\ell}}\right)\right]=\\
=Z_{N_{\ell}}^{*}C_{N_{\ell}}^{*}\left[\frac{1}{2^{N_{\ell}}}\sum_{\sigma_{V_{\ell}}\in\Omega^{V_{\ell}}}\exp\left[-\beta\lambda_{\ell}^{*}\bar{H}_{\ell}\left(\sigma_{Q_{\ell}}\right)\right]\right]^{\frac{1}{\lambda_{\ell}^{*}}}=\\
=\left[\frac{Z_{N_{\ell}}^{*}C_{N_{\ell}}^{*}}{2^{N_{\ell}/\lambda_{\ell}^{*}}}\right]\bar{Z}_{N_{\ell}}^{*}\left(\sigma_{Q_{\ell-1}}\right)^{\frac{1}{\lambda_{\ell}^{*}}}\label{eq:111}
\end{multline}
where $\lambda_{\ell}^{*}$ is the rate of the associated PPP. In
the last two steps we introduced the partition function $\bar{Z}_{N_{\ell}}^{*}\left(\sigma_{Q_{\ell-1}}\right)$
associated to the $\bar{H}_{\ell}$ model at slightly rescaled temperature
$\bar{\beta}^{*}=\lambda_{\ell}^{*}\beta$. Let $\bar{\xi}_{\ell}^{*}\left(\sigma_{Q_{\ell-1}}\right)$
the associated Gibbs measure associated to the $\bar{H}_{\ell}$ model
at rescaled temperature $\bar{\beta}^{*}$, then the formula for the
average becomes
\begin{multline}
f_{\ell-1}\left(\sigma_{Q_{\ell-1}}\right)=\sum_{\sigma_{V_{\ell}}\in\Omega^{V_{\ell}}}\xi_{\ell}\left(\sigma_{Q_{\ell}}\right)\,f_{\ell}\left(\sigma_{Q_{\ell}}\right)=\\
=\frac{1}{Z_{N_{\ell}}\left(\sigma_{Q_{\ell-1}}\right)}\sum_{\sigma_{V_{\ell}}\in\Omega^{V_{\ell}}}\exp\left[-\beta\,H_{\ell}\left(\sigma_{Q_{\ell}}\right)\right]f_{\ell}\left(\sigma_{Q_{\ell}}\right)=\\
=\frac{1}{Z_{N_{\ell}}\left(\sigma_{Q_{\ell-1}}\right)}\sum_{\sigma_{V_{\ell}}\in\Omega^{V_{\ell}}}\exp\left[-\beta_{\ell}H\left(\sigma_{V_{\ell}}\right)-\beta\bar{H}_{\ell}\left(\sigma_{Q_{\ell}}\right)\right]f_{\ell}\left(\sigma_{Q_{\ell}}\right)=\\
=\frac{Z_{N_{\ell}}^{*}}{Z_{N_{\ell}}\left(\sigma_{Q_{\ell-1}}\right)}\sum_{\sigma_{V_{\ell}}\in\Omega^{V_{\ell}}}\xi_{\ell}^{*}\left(\sigma_{V_{\ell}}\right)\exp\left[-\beta\bar{H}_{\ell}\left(\sigma_{Q_{\ell}}\right)\right]f_{\ell}\left(\sigma_{Q_{\ell}}\right)=\\
=\frac{Z_{N_{\ell}}^{*}C_{N_{\ell}}^{*}}{Z_{N_{\ell}}\left(\sigma_{Q_{\ell-1}}\right)}\left[\frac{1}{2^{N_{\ell}}}\sum_{\sigma_{V_{\ell}}\in\Omega^{V_{\ell}}}\exp\left[-\beta\lambda_{\ell}^{*}\bar{H}_{\ell}\left(\sigma_{Q_{\ell}}\right)\right]f_{\ell}\left(\sigma_{Q_{\ell}}\right)^{\lambda_{\ell}^{*}}\right]^{\frac{1}{\lambda_{\ell}^{*}}}=\\
=\left[\frac{Z_{N_{\ell}}^{*}C_{N_{\ell}}^{*}}{2^{N_{\ell}/\lambda_{\ell}^{*}}}\right]\frac{\bar{Z}_{N_{\ell}}^{*}\left(\sigma_{Q_{\ell-1}}\right)^{\frac{1}{\lambda_{\ell}^{*}}}}{Z_{N_{\ell}}\left(\sigma_{Q_{\ell-1}}\right)}\left[\sum_{\sigma_{V_{\ell}}\in\Omega^{V_{\ell}}}\bar{\xi}_{\ell}^{*}\left(\sigma_{Q_{\ell-1}}\right)f_{\ell}\left(\sigma_{Q_{\ell}}\right)^{\lambda_{\ell}^{*}}\right]^{\frac{1}{\lambda_{\ell}^{*}}}\label{eq:112}
\end{multline}
and putting together everything simplifies to
\begin{equation}
f_{\ell-1}\left(\sigma_{Q_{\ell-1}}\right)=\left[\sum_{\sigma_{V_{\ell}}\in\Omega^{V_{\ell}}}\bar{\xi}_{\ell}^{*}\left(\sigma_{Q_{\ell-1}}\right)f_{\ell}\left(\sigma_{Q_{\ell}}\right)^{\lambda_{\ell}^{*}}\right]^{\frac{1}{\lambda_{\ell}^{*}}}
\end{equation}
This formula does not depend from $N$ and then holds also in the
thermodynamic limit $N\rightarrow\infty$, where we can actually take
$\beta_{\ell}$ to zero, and applying to the recursion one would find
\begin{equation}
\lim_{\lambda_{\ell}^{*}\rightarrow1}\left[\sum_{\sigma_{V_{\ell}}\in\Omega^{V_{\ell}}}\bar{\xi}_{\ell}^{*}\left(\sigma_{Q_{\ell-1}}\right)f_{\ell}\left(\sigma_{Q_{\ell}}\right)^{\lambda_{\ell}^{*}}\right]^{\frac{1}{\lambda_{\ell}^{*}}}=\sum_{\sigma_{V_{\ell}}\in\Omega^{V_{\ell}}}\bar{\xi}_{\ell}\left(\sigma_{Q_{\ell-1}}\right)f_{\ell}\left(\sigma_{Q_{\ell}}\right)
\end{equation}
The idea is that for positive $f_{\ell}$ and in the limit $q_{\ell}-q_{\ell-1}\rightarrow0$
one would have that the SK average is taken at infinite temperature,
then equivalent to a PPP of rate $\lambda_{\ell}^{*}\rightarrow1$.

\section{Conclusive remarks}

\begin{figure}
\begin{centering}
\includegraphics[scale=0.2]{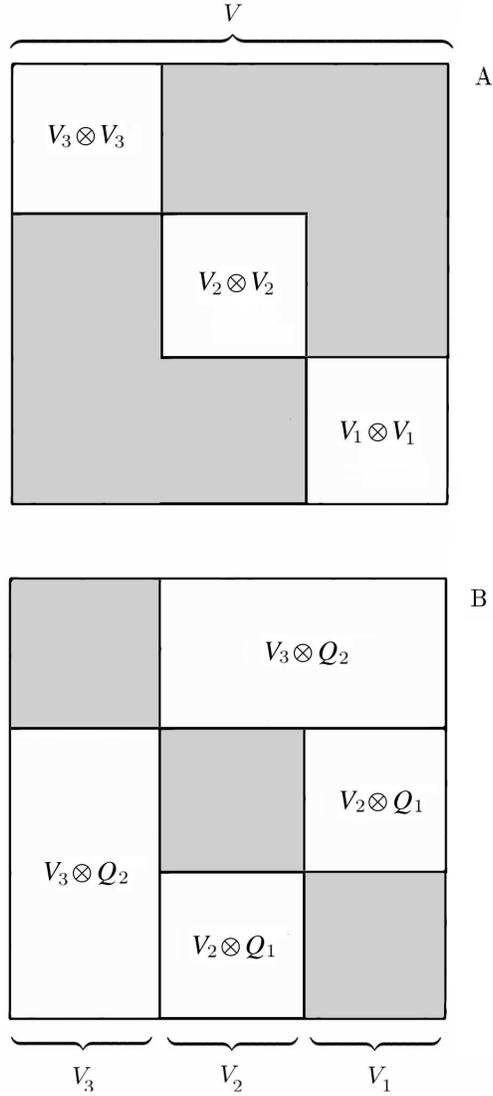}
\par\end{centering}

\raggedleft{}~\caption{\label{fig:Partition-of--2}Two extreme pictures for the RSB ansatz
for $L=3$. The diagram shows the edges that contributes to the energy
in the orthodox MF ansatz, top figure A, where the Hamiltonian operator
is diagonal under Gibbs measure, and a situation where the interfaces
dominate the total energy, lower figure B. We expect the second option
to be much more likely for fully connected models, because in such
models the interfaces are overwhelmingly large respect to the contribution
from edges between spins of the same layer.}
\end{figure}

Even if we easily obtained the functional, from the physical point
of view this short analysis still didn't clarified what is the proper
approximation for $\Delta\bar{H}\left(\sigma_{Q_{\ell}}\right)$ under
the full measure $\mu$ (see Figure \ref{fig:Partition-of--2}). If
one assume that the same approximation used under $\xi_{\ell}$ holds
also under $\mu$ it would be equivalent to assert that 
\begin{equation}
H_{\ell}\left(\sigma_{Q_{\ell}}\right)\rightarrow\sqrt{q_{\ell}-q_{\ell-1}}\,H\left(\sigma_{V_{\ell}}\right)
\end{equation}
and then the model would be simply a sum of smaller independent systems
at higher temperatures. By the way, we remark once again that for
large $L$ the coefficients $q_{\ell}-q_{\ell-1}$ vanish respect
to $q_{\ell-1}$, and is unlikely that this ansatz can return stable
solutions in any fully connected model. 

In fact, this would be a quite orthodox mean-field ansatz \cite{Opper Saad}
where the external field acting on the layer is irrelevant, although
in SK the number of pairwise energy contributions from the interfaces
is much larger than the energy contributions from the spins in the
same layer, as already predicted in \cite{Franz}. We expect that
the proper approximation under $\mu$ would be the Generalized Random
Energy model (GREM) \cite{Bolt,PaK}, where 
\begin{equation}
H_{\ell}\left(\sigma_{Q_{\ell}}\right)\stackrel{d}{=}\epsilon_{\ell}\left(N\right)+\sqrt{N}\,\bar{\gamma}_{\ell}g_{\ell}\left(\sigma_{Q_{\ell}}\right)
\end{equation}
and $\lambda_{\ell}$ is the sequence of free parameters that controls
the variance, and $g_{\ell}\left(\sigma_{Q_{\ell}}\right)$ is a collection
of normal random variables of covariance matrix 
\begin{equation}
E\left[g_{\ell}\left(\sigma_{Q_{\ell}}\right)g_{\ell}\left(\tau_{Q_{\ell}}\right)\right]=\prod_{i\in Q_{\ell}}I\left(\sigma_{i}=\tau_{i}\right),
\end{equation}
with $E\left(\,\cdot\,\right)$ representing the normal average that
acts on the variables $g_{\ell}\left(\sigma_{Q_{\ell}}\right)$. The
difference with the orthodox MF ansatz is in that by changing $g_{\ell}^{*}\left(\sigma_{V_{\ell}}\right)$
with $g_{\ell}\left(\sigma_{Q_{\ell}}\right)$ for any magnetization
states $\sigma_{V}$ and $\tau_{V}$ with $\sigma_{V_{\ell}}=\tau_{V_{\ell}}$,
$\sigma_{Q_{\ell-1}}\neq\tau_{Q_{\ell-1}}$ now one has 
\begin{equation}
E\left[H_{\ell}\left(\sigma_{Q_{\ell}}\right)H_{\ell}\left(\tau_{Q_{\ell}}\right)\right]=0
\end{equation}
instead of $\bar{\gamma}_{\ell}^{2}N$. This computing scheme is essentially
a Guerra-Toninelli interpolation between the layers, a method first
used by Billoire \cite{Sbirulen} to compute the finite size corrections
to the SK model. One can easily verify that both ansatz gives the
same recursive formula for the average, but this ansatz, that we interpret
as fully equivalent to the RSB ansatz, bases on the fact that for
large $L$ the layer behavior is mostly dominated by the interface
interaction from the previous layers,
\begin{equation}
H_{\ell}\left(\sigma_{Q_{\ell}}\right)\rightarrow\sqrt{q_{\ell-1}}\,\sigma_{V_{\ell}}\cdot h_{V_{\ell}}\left(\sigma_{Q_{\ell-1}}\right),
\end{equation}
which seems the case indeed for any fully connected mean-field model,
at least. Notice that in the termodynamic limit the associated Gibbs
measure is distributed proportionally to a cascade of PPP, known as
Ruelle Cascade \cite{PaK,Bolt,ASS,kUltrapanchenko}, that is known
to have an ultrametric overlap support. For SK this property has been
first proven in \cite{kUltrapanchenko}, where it is shown that the
Gibbs measure of the SK model can be infinitesimally perturbed into
a Ruelle Cascade. 

In conclusion, it seems not possible to distinguish between the orthodox
mean field ansatz (the Gibbs measure is a product measure) from the
RSB ansatz (the measure is a Ruelle Cascade) by only looking at the
Parisi Formula. Nonetheless, we argument that the orthodox mean field
theory is unlikely to hold in SK, due to expected dominance of the
interface contribution. Weather an orthodox mean-field ansatz is meaningful
in some sense for the SK model we still cannot say, although it seems
related to the replica trick. Despite this, we think it would naturally
apply to many other disordered systems, like random polymers, or any
other model with low connectivity between the layers.

\section{Acknowledgments}

I wish to thank Amin Coja-Oghlan (Goethe University, Frankfurt) for
sharing his views on Graph Theory and Replica Symmetry Breaking, which
deeply influenced this work. I also wish to thank Giorgio Parisi,
Pan Liming, Francesco Guerra, Pietro Caputo, Nicola Kistler, Demian
Battaglia, Francesco Concetti and Riccardo Balzan for interesting
discussions and suggestions. This project has received funding from
the European Research Council (ERC) under the European Union’s Horizon
2020 research and innovation programme (grant agreement No {[}694925{]}).

\end{document}